# GUIDELINES TO MINIMIZE THE COST OF SOFTWARE QUALITY IN AGILE SCRUM PROCESS

Deepa Vijay[1] and Gopinath Ganapathy[2]

Department of Computer Science & Engineering, Bharathidasan University, Trichy, India

## *ABSTRACT*

*This paper presents a case study of Agile Scrum process followed in Retail Domain project. This paper also reveals the impacts of Cost of Software Quality, when agile scrum process is not followed efficiently. While analyzing the case study, the gaps were found and guidelines for process improvements were also suggested in this paper.*

## 1.INTRODUCTION

Agile software development approaches have become increasingly popular during the last few years. Agile methodology is an alternative to traditional project management now widely used in software development to increase the project quality, reduce time to market and to increase customer satisfaction.

Agile methods have gained tremendous acceptance in the commercial arena since late 90s because they accommodate volatile requirements, focus on collaboration between developers and customers, and support early product delivery. Two of the most significant characteristics of the agile approaches are:1) they can handle unstable requirements throughout the development lifecycle 2) they deliver products in shorter timeframes and under budget constraints when compared with traditional development methods [1]. Many reports [Agile Project Success Rate survey results] support the advantages of agile methods [10].

This paper details a case study of Retail domain project which failed to adopt agile scrum process as per agile standards. The consequences faced due to improper adoption of agile process are documented. The rest of the paper is organized as follows:

Section 2 presents a Literature Survey of Agile methodology, section 3 presents a case study of an agile scrum project, Guidelines are described in section 4 and section 5 is a conclusion.

## 2. LITERATURE SURVEY

Research has been undertaken into what is meant by agility and different types of agile methods and how a supposed agile method should be adopted in a project.

Agile software development is a group of software development methods based on iterative and incremental development, where requirements and solutions evolve through collaboration between self-organizing, cross-functional teams. It promotes adaptive planning, evolutionary development and delivery, a time-boxed iterative approach, and encourages rapid and flexible response to change. It is a conceptual framework that promotes foreseen interactions throughout the development cycle [2]. One of the fastest-growing agile methods is Scrum. It was formalized over a decade ago by Ken Schwaber and Dr. Jeff Sutherland, and it's now widely being used by large and small companies [4].

Well-known agile software development methods include [6]:

• Extreme Programming
• Crystal methodologies family
• Feature-Driven Development
• Agile Modeling
• Scrum

### 2.1. Extreme Programming (XP)

Extreme Programming was introduced by Kent Beck in 2000.     Extreme Programming is one of several popular Agile Processes [11] XP is a lightweight, efficient, low-risk, flexible, predictable, scientific, and fun way to develop software. [12]. It also promotes customer involvement as proposed in agile where customer is utmost priority. It promotes continuous testing, continuous feedback and planning and close teamwork to deliver working software at frequent intervals. In XP customer works with development team also referred to as "User Stories". The development team delivers highest priority tested working software on iteration by iteration basis.

### 2.2. Crystal Methodology

Crystal methodology is most lightweight approaches of software development. Crystal is comprised of family of agile methodologies such as crystal clear, crystal yellow, crystal orange whose unique characteristics are governed by team size and project priorities. Each of the Crystal methodologies requires certain roles, policy standards, products and tools to be adopted. Crystal Clear, which is one of the Crystal methodologies, can be applied to development teams of six to eight members, working on non-life critical systems. It focuses on people, not processes of artifacts [13].

### 2.3. Feature Driven Development

Feature Driven Development (FDD) was founded by Jeff De Luca and Peter Coad. It combines some practices recognized in the industry into one methodology. These practices are all determined from a client valued functionality (feature) viewpoint. As of other agile methodologies, its key goal is to deliver tangible, working software repeatedly in a timely manner [14].

### 2.4. Agile Modeling

Agile Modeling (AM) is a methodology for modeling and documenting software systems based on best practices. It is a collection of values and principles that can be applied on an (agile) software development project. This methodology is more flexible than traditional modeling methods, making it a better fit in a fast changing environment. It is part of the Agile software development tool kit [15].

Agile Modeling is a supplement to other agile methodologies such as Scrum, Extreme Programming (XP), and Rational Unified Process (RUP).

## 2.5. Scrum

Scrum is a lightweight process framework for agile development, and the most widely-used one.[16].

- A "process framework" is a particular set of practices that must be followed in order for a process to be consistent with the framework. (For example, the Scrum process framework requires the use of development cycles called Sprints, the XP framework requires pair programming, and so forth.)
- "Lightweight" means that the overhead of the process is kept as small as possible, to maximize the amount of productive time available for getting useful work done.

### 2.5.1. Understanding of Scrum Framework

Scrum is an iterative and incremental agile software development framework for managing software projects and product or application development. Its focus is on "a flexible, holistic product development strategy where a development team works as a unit to reach a common goal" as opposed to a "traditional, sequential approach"[3]

The Scrum framework consists of Scrum Teams and their associated roles, events, artifacts, and rules. Each component within the framework serves a specific purpose and is essential to Scrum's success and usage. The rules of Scrum bind together the events, roles, and artifacts, governing the relationships and interaction between them. [4]

The Scrum teams have two management roles. Scrum Master who manages the process and a Product Owner who manages the product. The third role, simply The Team, does the work. The Team is a cross functional team which may include analysts, programmers, UI developers/designers, Business Analyst and quality assurance engineers. The Product owner is an individual who facilitates knowledge sharing between domains. One important rule in Scrum is to time-box activities [8]. Scrum development is organized into time-boxed iterations, called sprints, which normally last 2 –4 weeks. Scrum makes extensive uses of structured lists of tasks:
The Product Backlog is a prioritized list of requirements needed in the product. Requirements need not be precise nor do they need to be described fully. As with most projects, the requirements are sourced from the expected users or "the business". The Product Owner prioritizes the Product Backlog: items of importance to the project/business, i.e. those items that add immediate and significant business value, are bubbled up to the top.

Sprint Backlog is a list of tasks to turn the Product Backlog for one Sprint into an increment of potentially shippable product. A burndown is a measure of remaining backlog over time. The Release Burndown measures remaining Product Backlog across the time of a release plan. A Sprint Burndown measures remaining Sprint Backlog items across the time of a Sprint.

Each sprint comprises development time punctuated by meetings. During the planning meeting, which is held immediately before the each sprint, the product owner selects tasks from the product backlog while team members estimate the effort required for each task and self-determine who will complete each task. Each day of development starts with a time-boxed (15 minute) daily scrum meeting in which team members describe what they have done since the previous Scrum, what they are going to do next and any obstacles faced. Daily scrums facilitate development of shared mental models, which are crucial for team coordination. Scrum also prescribes review meetings where progress is reviewed and retrospective meeting, where the team discusses the process and practices used, if needed; make changes to their way of working before the next sprint starts. Fig.1 shows the Scrum Process

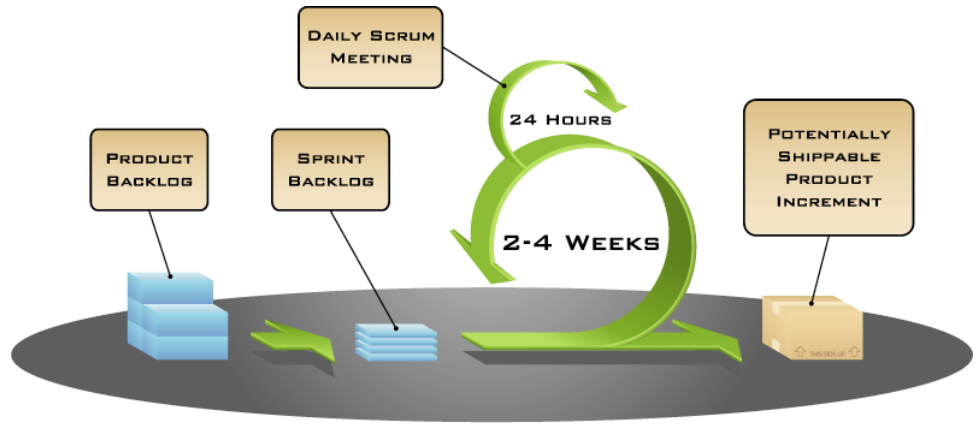

Fig. 1. Scrum Framework

## 3.EMPRICAL CASE STUDY

A study has been done in one of the modules (ABC) of Retail domain project (XYZ) in a software company which followed agile scrum framework for over 3 years. The project offers travel services and followed on Onsite-Offshore model. 3 peoples including Product Owner sits at Onsite and 13 peoples at offshore. The team followed agile scrum framework as described in Fig 1. Sprint cycle is lasted for 2 weeks. All the requirements are placed in product backlog and maintained by product owner. Sprint planning meeting is scheduled before every sprint. Only Onsite team members were part of sprint planning meeting as the meeting happened during onsite timings. Product owner prioritizes the sprint items and offshore team provides estimates for each user stories. Once the sprint backlog is ready, team starts working on sprint items. Envisioning, design and development and testing activities were carried out throughout the sprint. Every day, daily scrum meeting was scheduled between onsite and offshore team members through telephone and discussed what has been done? What will be done? And any blockers? This meeting lasts for 15 min to 30 mins. At the end of each iterating sprint demo is given to the customers for their feedback and also Retrospective meetings are scheduled to discuss what went well? What didn't go well? Lessons learned and improvements. Even though the project adopted agile scrum framework, below are the negative impacts observed.

(1)  Teams were overloaded with sprint backlog items and unable to complete the planned items
(2)  Team members were need to stretch back and unable to deliver quality product due to stringent time lines and stress
(3) Cost of Quality is high
(4) More defects appear at the end and Last iterations produce less new functionality
(5) Re-factor and re-design cost too much and take too long
(6) End product is not as per the customer requirements and expectation

(7) Delayed in new feature design and delivery
(8) More number of production defects
(9) Rework is more
(10) Loss of reputation
(11) Loss of customer

A thorough study has been done on the above project and resulted in finding the gaps in adopting agile scrum process. Below Table. 1 explains the gaps identified in adopting of scrum process.

| Scrum Practice | Gaps in adopting scrum process |
| --- | --- |
| Envisioning Meeting | It was observed that the technical and architectural Envisioning meetings were not effective. Since all the envisioning meetings happens through telecall. Interpretation and understanding of the feature is different with respect to Dev, QE and Product |
| Sprint Planning Meeting | Most of the tasks were completed by offshore teams but, they never been a part of Sprint planning meetings due to Time Zone difference. This created a huge gap and kept offshore teams overloaded all the time with sprint items. Due to this, few development process like code reviews, unit testing and code refactoring are skipped and also Q team members never get sufficient time to review test cases, perform impact analysis and many times automation process was skipped. |
| Sprint Backlog | The moment Sprint backlog is identified, immediately sprint backlog should be freezed for that sprint. Since the sprint items were not freezed and requirements were getting added continuously throughout the sprint. It is difficult for any dev and QE teams to incorporate change requirements and also observed that most of the time is spent only in reworking of the changed requirements than spending time in new feature development and testing |
| User Story | Most of the user stories are Poorly defined [8]. A story should be INVEST (*Independent, Negotiable, Valuable, Estimable, Small, Testable*). Business justification are missing and also acceptance criteria were also missing |
| Definition of DONE | Lack of team understanding of the Done criteria<br>A feature is considered as DONE when it is completed by the entire team and ready to be shipped. But the gap is observed that, once the coding is done, it is considered as DONE from the team irrespective whether story is tested or not. |

International Journal of Software Engineering & Applications (IJSEA), Vol.5, No.3, May 2014

| | |
|---|---|
| Sprint Demo | End of each sprint, Sprint Demo is shown to the customers by the team members. The aim of this demo is to present the task completed in particular sprint and get the feedback from the customers. Most of the time it is observed that team have not got any feedback until the feature is released to production. So Sprint demo is not effective. Customers participation is very less |
| Sprint Retrospective Meeting | Sprint Retrospective meetings were held after each sprint to discuss good, bad and improvements. It is observed that the output from these meetings was not yet utilized effectively. The action points were not analyzed and implemented. |
| Requirements Meeting | Most of the time it is observed that requirement clarification is done through telephonic discussion. Team members are failed to record the discussion in respective user story in bug tracking tool. This leads to increase in turnaround time asking clarification with Dev team and with product |

Table 1. Process Gaps

### 3.1. Data Gathering and Analysis

The objective of data collection is to find the phase where more number of defects are leaked to production and perform Root Cause Analysis of the defects. Collected production defects for the year 2012 and performed RCA. Table 2. Shows the defect count and reasons for defect leakage.

Table 2. Defect Count and Root Cause Reasons

| No. of Defects | Root Cause Category |
|---|---|
| 36 | Missed in Envisioning |
| 5 | Missing Component Test Coverage |
| 3 | Missed in Integration Test Coverage |
| 44 | Missing Unit Test Coverage |
| 2 | Missing Usability Test Coverage |
| 6 | Test Environment Limitation |
| 45 | Others |

Source: This is real time data gathered from Retail domain project

From Table 2 it is clear that 25% of the defects are leaked due to Missed in Envisioning (Escaped in Envisioning activity) and 31% of the defects were leaked due to Missing Unit Test Coverage.

International Journal of Software Engineering & Applications (IJSEA), Vol.5, No.3, May 2014

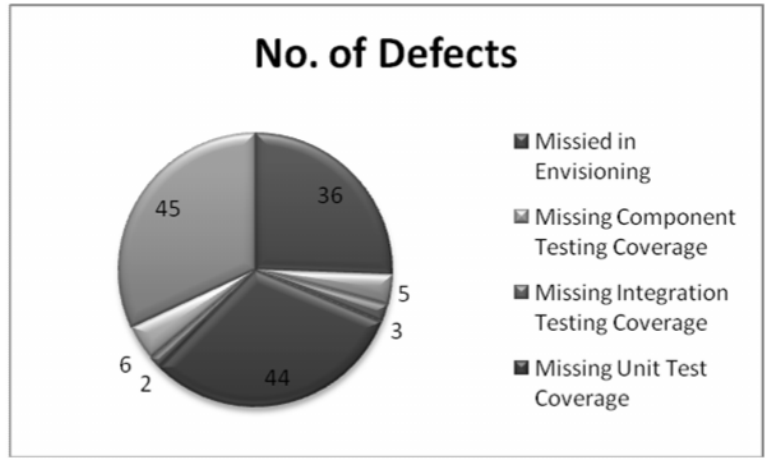

Fig. 2. RCA of Production Defects

### 3.2. Cost Calculation

Cost of the project has been calculated as per the below Table 3. Assuming that Cost of the defect found in requirements phase is $1 and cost of the defect found in Design/Development phase is $10 and costs $100 if found in Testing and $1000 if it is found in Production environment.

Table 3. Cost Incurred in Each Phase

| Phase : | Requirements Envisioning | Design/Unit Testing | Integration Testing | Production |
|---|---|---|---|---|
| Defects found at : | Own Process Step | Next Process Step | Later Process Step | Customer |
| Cost | $1 | $10 | $100 | $1000 |
| Impact | Minor | Minor Delay | Significant Rework and Delay in delivery of the feature | Rework Warranty Cost Loss of Reputation |
|  |  |  |  | 90 |
|  |  |  |  | **$90000** |
| Defects could have found earlier at : | 36 | 44 | 10 |  |
| Cost | $36 | $440 | $1000 | $1476 |

In this case study, Envisioning and Testing phase defects were considered for cost calculation. From the above Table, it is clear that the bugs found in post production environment are high because the scrum process was not followed efficiently as per the standards by the team. Cost of
67

rework is high which in turn increase in cost of software quality. If the bugs would have found in earlier phases, cost of rework would have been minimized.

## 4. GUIDELINES

1. As we know that, there are certain gaps in successful envisioning process, due to the gaps, most of the time the envisioning meetings are not very effective. To make it effective, it is recommended to all the team members to have a Pre-envisioning phase before final envisioning meeting.

**Pre-Envisioning:** Team members should be given a high level description of the new feature/enhancement. Team members should understand the feature and come up with scenario checklist/questionnaires/new ideas before commencement of the final envisioning meeting. All the checklist/questionnaires/new ideas of all the team members are discussed in final envisioning meeting. The idea of having pre-envisioning meeting is to bring common knowledge about the new feature/enhancement and minimal chances of missing critical scenarios.

2. Time Zone difference is one of the main reasons to skip Sprint Planning Meetings. This gap can be eliminated with minimal effort and easy way just organizing the sprint planning meeting in a common time zone.

If there are less chances of implementing sprint planning meeting in a common time zone, this gap can be handled by requesting the product owner to share the priority list of items for the upcoming sprint to the offshore team members and ask them to estimate the same. Based on the offshore team estimation, the items for the upcoming sprint should be finalized. By implementing this process change, there are less chances of skipping quality activities like reviews, unit testing etc.

3. Sprint backlog should be frozen for the current sprint unless there is a critical requirement/bug should be addressed immediately.

4. As part of Sprint Planning activity, team members are expected to understand the user stories thoroughly before estimation. They should also check each and every story planned for the print to ensure that business justification & acceptance criteria are met.

5. The feature is said to be done only when it can be ready to be shipped.

6. Retrospective meeting is a very important activity. Make sure that every team member should get involved and list down the important action points. Based on the priority, team members should start working on the action points to provide right solution.

The above guidelines are very simple and easy to implement. Even though these guidelines look simple, it can bring huge difference in output.

## 5. CONCLUSIONS

In this paper a through case study was done using real time data of an agile team, which failed to follow scrum process effectively. Cost of rework is measured as part of the study. Gaps were were identified while following Agile Scrum process, and simple guidelines are suggested to improve the process. If these guidelines are followed, the cost of software quality can be minimized in Agile Scrum project.

## REFERENCES

[1]   Ming Huo, June Verner, Liming Zhu, Muhammad Alibabar "Software Quality and Agile Methods", In Proceedings of Annual International Computer Software and Applications Conference (COMPSAC'04).
[2]   Agile Manifesto [Online] Retrieved 14 June 2010. Available at: http://agilemanifesto.org/

[3] Agile Software Development [Online] Available at: http://en.wikipedia.org/wiki/Agile_Software_Development
[4] Ken Schwaber and Jeff Sutherland (2011). "The Definitive Guide to Scrum: The Rules of the Game"
[5] http://www.associatedcontent.com/article/614143/software_quality_assurance_agile_testing.html
[6] http://en.wikipedia.org/wiki/Agile_software_development
[7] Ronen Bar-Nahor , Why Scrum projects might fail?
[8] Ralph, Paul and Shportun, Petr, "Scrum Abandonment in Distributed Teams: A Revelatory Case" (2013). Pacific Asia Conference on Information Systems (PACIS), 2013.
[9] Malik F. Saleh ," An Agile Software Development Framework", International Journal on Software Engineering (IJSE), Volume (2): Issue (5): 2011
[10] Agile Success Survey Results [Online] Available at : http://www.ambysoft.com/surveys/agileSuccess2010.html
[11] Extreme Programming [Online] Available at : http://www.extremeprogramming.org
[12] Kent . B, "Extreme Programming explained", 2nd edition, Addison- Wesley, Boston, 2004
[13] Crystal Clear (software development) [Online] Available at: http://en.wikipedia.org/wiki/Crystal_Clear_(software_development)
[14] Feature Driven Development. [Online] Available at: http://en.wikipedia.org/wiki/Feature_Driven_Development
[15] Agile Modeling. [Online] Available at: http://en.wikipedia.org/wiki/Agile_Modeling
[16] https://www.cprime.com/resources/what-is-agile-what-is-scrum/

**Authors**

**Deepa Vijay,** Bangalore, India. B.E.S. Electronics Science, Madras University, India, M.Sc. Computer Technology, Periyar University, Salem, India, M.Phil. Computer Science, Alagappa University, India. Ph.D scholar in Department of Computer Science, Bharathidasan University, Trichy, India.

**Dr. Gopinath Ganapathy, Trichy, India**. B.Sc. Computer Science St.Joseph's College Bharathidasan University,Trichy,India, Masters in Computer Applications, St.Joseph's College Autonomous, Trichy, India Ph.D, Madurai Kamaraj University, India.He is currently the Chair and Head School of Computer Science & Engineering, Bharathidasan University, Trichy, India.Dr. Gopinath Ganapathy is a Professional Member in IEEE, Professional Member in ACM (USA) and Life member in Computer Society of India (CSI).